\documentclass[aps,prl,reprint,twocolumn,letterpaper,floatfix,preprintnumbers,superscriptaddress,groupedaddress]{revtex4-2}
  
\usepackage{graphicx}
\usepackage{dcolumn}
\usepackage{bm}
\usepackage[dvipsnames]{xcolor}
\usepackage{color}
\usepackage{epsfig}
\usepackage{latexsym}
\usepackage{amsmath}
\usepackage{amsfonts}
\usepackage{rotating}
\usepackage[normalem]{ulem}

\providecommand{\ie}{i.e.,}  

\begin{document}

\title{$Z_2$ topology of bismuth}

\author{Irene Aguilera}
\email[Corresponding author: ]{i.aguilera@fz-juelich.de}
\altaffiliation[Current affiliation: ]{IEK5-Photovoltaik, Forschungszentrum J{\"{u}}lich, 52425 J{\"{u}}lich, Germany.}
\author{Hyun-Jung Kim}
\author{Christoph Friedrich}
\author{Gustav Bihlmayer}
\author{Stefan Bl\"{u}gel}
\affiliation{Peter Gr{\"{u}}nberg Institute and Institute for Advanced Simulation, Forschungszentrum J{\"{u}}lich and JARA, 52425 J{\"{u}}lich, Germany}

\date{\today}

\begin{abstract}
While first-principles calculations with different levels of sophistication predict a topologically trivial $Z_2$ state for bulk bismuth, some photoemission experiments show surface states consistent with the interpretation of bismuth being in a topologically non-trivial $Z_2$ state. We resolve this contradiction between theory and experiment by showing, based on quasiparticle self-consistent $GW$ calculations, that the experimental surface states interpreted as supporting a non-trivial phase are actually consistent with a trivial $Z_2$ invariant. We identify this contradiction as the result of a \emph{crosstalk} effect arising from the extreme penetration depth of the surface states into the  bulk of Bi. 
A film of Bi can be considered bulk-like only for thicknesses of about 1000 bilayers ($\approx$ 400~nm) and more.
\end{abstract}

\maketitle

The topological classification of crystals is a very powerful and successful theoretical concept. It underlies, for example, the bulk-boundary correspondence, which relates the bulk topological invariant to the number of robust chiral edge or surface states localized at the boundaries of the material. 
Hence, experimental efforts have been dedicated to observing boundary localized gapless states to validate their topological character~\cite{bernevig2006_2,hsieh2008,Drozdov2014,Zhang2010_2}. Along with the development of topological band theory, right from the beginning, Bi has been regarded as a prototypical material. The Bi$_{1-x}$Sb$_x$ alloys were among the first materials theoretically predicted~\cite{fu2007-2}, and the first experimentally detected, as non-trivial topological insulators in three dimensions~\cite{hsieh2008}. Elemental Bi, however, was known for many years simply as a trivial semimetal in terms of a $Z_2$ classification~\cite{kane2005-1}. Therefore, no time-reversal symmetry protected surface states were expected. On the other hand, a space-group symmetry based $Z_4$ classification unveiled a non-trivial topology, suggesting that bulk Bi presents a topological crystalline insulator phase~\cite{hsu2019} and a higher-order topological phase~\cite{schindler2018} that supports one-dimensional hinge states in a rotational-symmetry preserving rod. 

\begin{figure}[!b]
\begin{center}
\includegraphics[angle=0,width=8.2cm]{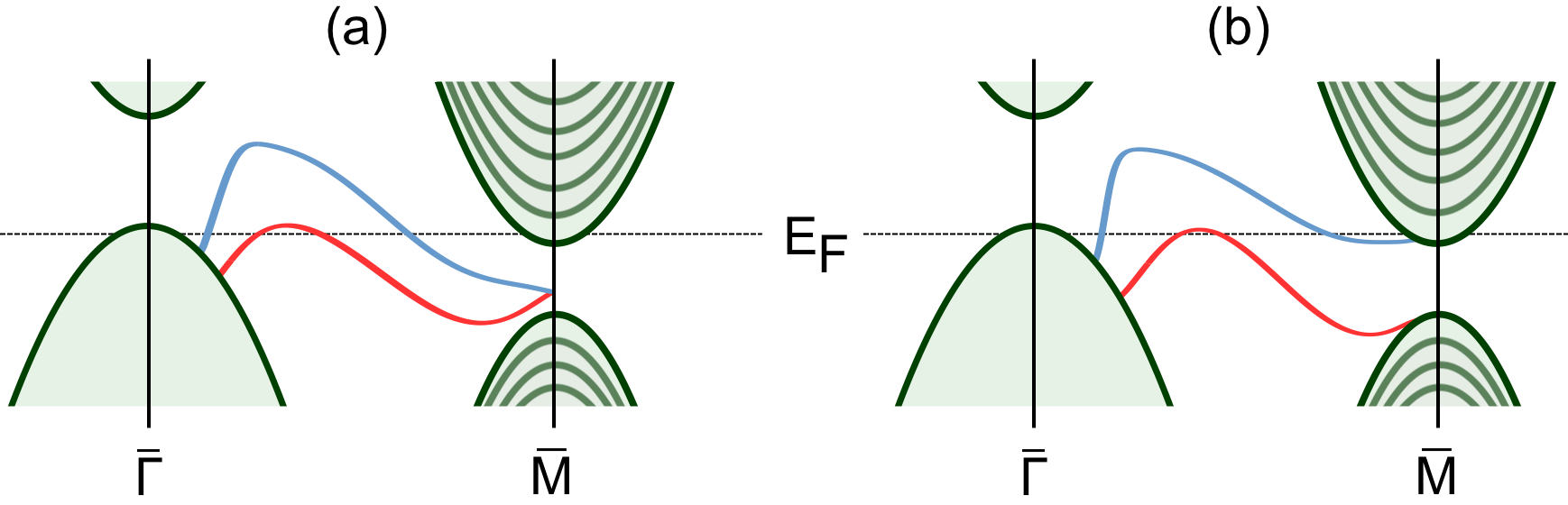}
\caption{\label{figscheme}  The two controversial behaviors of the surface states of Bi films in the $\bar{\rm{\Gamma}}-\bar{\rm{M}}$ direction: (a) trivial surface states as predicted by DFT calculations and (b) experimentally observed connectivity of the surface states with the bulk at the $\overline{\rm{M}}$ point.}
\end{center}
\end{figure}

However, experimentally, the interpretation of Bi being in a topologically trivial $Z_2$ state  was challenged by angle-resolved photoemission spectroscopy (ARPES) used  to investigate the dispersion of the surface states in the $\overline{\Gamma}-\overline{\rm{M}}$ direction of the Brillouin zone. In 2013, an experimental work~\cite{ohtsubo2013} claimed a non-trivial $Z_2$ topology for bulk Bi. In 2016, this work was followed by several other experiments~\cite{ohtsubo2016,yao2016,ito2016} that also displayed 
surface states that connect to bulk valence and conduction bands consistent with a non-trivial interpretation [as shown in Fig.~\ref{figscheme}(b)]. Thus, the determination of the $Z_2$ topological character of bismuth has been elusive. The reason for this is the tiny ($\sim$11$-$15~meV) direct band gap of the Bi band structure at the L point of the bulk Brillouin zone, which projects onto the $\overline{\rm{M}}$ point of the surface Brillouin zone. An inversion of the valence and conduction bands at this $\mathbf{k}$ point would change the topological order of Bi. In other words, it is the symmetry (parity) of the wavefunctions of the highest valence band at the L point what determines the $Z_2$ topological character of Bi.

First-principles calculations of bulk Bi, even when carried out with very different theoretical methods~\cite{aguilera2015,hsu2019}, tend to show that the electronic structure of Bi is described by a trivial $Z_2$ invariant,
and density functional theory (DFT) calculations of films of Bi(111) display trivial surface states~\cite{koroteev2004,zhang2009_2,chang2019}, 
similar to those shown schematically in Fig.~\ref{figscheme}(a). 
The two surface states (red and blue), that are a Kramers pair, are degenerate at the 
time-reversal invariant momentum $\overline{\rm{M}}$ of the two-dimensional surface Brillouin zone. The obvious contradiction between experiment and theory has remained unresolved to this day and is a result of a missing understanding.

The discrepancies between theory and experiment have been attributed
to the failure of DFT to predict the size of band gaps correctly. This is not entirely improbable given the very small band gap of bulk Bi at the L point.
If one improves the DFT calculation by including many-body effects using the quasiparticle self-consistent $GW$ (QS$GW$) method~\cite{faleev2004} and a consistent treatment of spin-orbit coupling, the band gap indeed decreases, but it does not change sign~\cite{aguilera2015}, thus again supporting the trivial character of bulk Bi~\footnote{Of course, there is still the possibility that high-order self-energy diagrams may eventually flip the sign, but we show here that there is no need for more accurate calculations to reconcile the literature.}.

Several theoretical works have focused on trying to resolve the contradiction between theory and experiment by searching for non-trivial $Z_2$ phases of Bi: improving the theoretical approximations~\cite{aguilera2015,koenig2021}, adding strain~\cite{aguilera2015,chang2019,koenig2021}, or doping~\cite{jin2020}. But these works seem to disregard that there are also experiments, not based on photoemission, that support the theoretical predictions of a trivial $Z_2$. For example, Kang \emph{et al.}~\cite{kang2021} have recently shown with angle-dependent magnetoresistance experiments that Bi and Sb exhibit markedly different behavior due to the different parities of the highest valence band at the L point. In other words, Bi and Sb must have a different $Z_2$ topological invariant (trivial and non-trivial, respectively). Thus, it is not sufficient to address only the contradiction between theory and ARPES, because the confusion around the $Z_2$ character of Bi goes beyond that.

In this Letter, we consider this issue from a more fundamental perspective that reconciles the very different conclusions of the experimental and theoretical literature. For that, we explore with QS$GW$ calculations both finite films of Bi and the thermodynamic limit by comparison with semi-infinite calculations.
In this way  
we intend to provide a consistent picture of the 
electronic structure of bulk and thin films of Bi where the seemingly contradictory theoretical and experimental results in the literature actually reconcile and both are consistent with a trivial $Z_2$ topology. 

Following the method described in Ref.~\cite{aguilera2019}, we parameterize a tight-binding (TB) Hamiltonian that 
accurately reproduces the \emph{bulk} QS$GW$ band structure of Ref.~\cite{aguilera2015}. (The reader is referred to 
this reference for the convergence parameters of the QS$GW$ calculation.) The accuracy of the method is shown for DFT in Fig.~S1(a) of the Supplemental Material~\cite{suppmat2021}. The interpolated QS$GW$ bulk band structure is shown in Fig.~S1(b). The TB Hamiltonian is expressed in a basis of Wannier functions~\cite{wannier90} and, thus, calculated fully \textit{ab initio} 
with the QS$GW$ method without the need for adjustable parameters to fit experimental results. 
We subsequently use the tight-binding parameters obtained for the bulk to construct the QS$GW$-TB Hamiltonian of films of bismuth of different thicknesses.
 
\begin{figure}
\begin{center}
\includegraphics[angle=0,width=8cm]{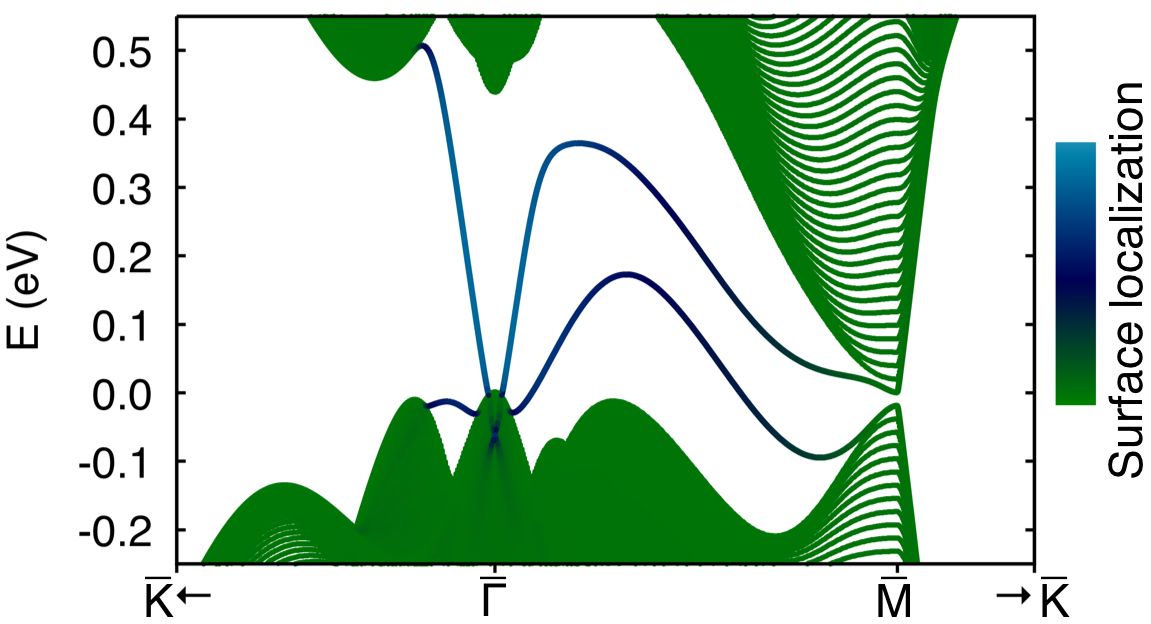}
\caption{\label{fig2-200} Band structure of a 200-BL film of Bi(111) along some high-symmetry lines in the two-dimensional Brillouin zone obtained with the QS$GW$-TB method. The color scheme represents the degree of localization of a state at  the surface, with blue representing surface states and surface resonances in the band gap of the film-projected bulk states, and green representing the bulk-like character.} 
\end{center}
\end{figure}

The band structure of a film of Bi(111) of 200-bilayer (BL) (approximately 80~nm) is shown in Fig.~\ref{fig2-200}. Unexpectedly, the resulting surface states do not look like in the previous DFT calculations, with the two states touching at the  $\overline{\rm{M}}$ point, but instead, 
they look very similar to the experimental ones (see Fig.~3(b) of Ref.~\cite{ito2016} for a 202 BL sample measured with ARPES). 
Given this result, the inconsistency between theory and experiment seems to be resolved. 
The surface states in Fig.~\ref{fig2-200} display indeed a connectivity to the bulk states that 
suggests a non-trivial topology [in the manner of Fig.~\ref{figscheme}(b)]. This raises the question whether the surface states of this film are topologically protected, even though the bulk is trivial.

\begin{figure}
\begin{center}
\includegraphics[angle=0,width=7.6cm]{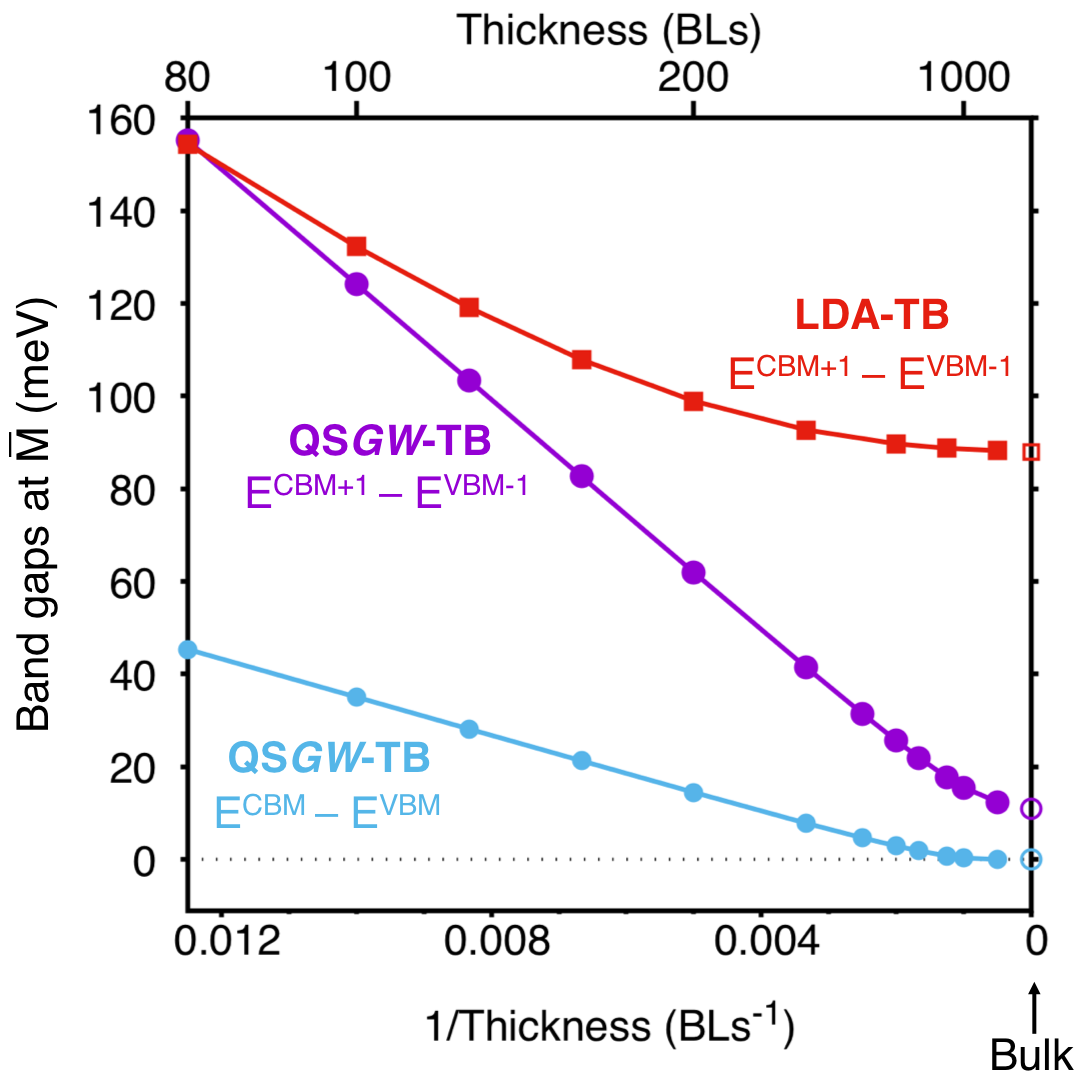}
\caption{Band gap at $\overline{\rm{M}}$ as a function of the thickness (top horizontal axis) and its inverse (bottom horizontal axis) of the Bi films in units of bilayers. Red squares and big purple circles represent the energy difference between the second conduction band and the last but one valence band obtained with TB parameters from DFT (LDA) and QS$GW$ calculations, respectively. The QS$GW$-TB results for the actual band gap at $\overline{\rm{M}}$ is shown by cyan small circles. The open symbols represent the bulk values corresponding to infinite thickness (or zero at inverse thickness).\label{fig3-thick}}
\end{center}
\end{figure}

This question is addressed by performing calculations of the band gap at $\overline{\rm{M}}$. We show this gap in Fig.~\ref{fig3-thick} as a function of the thickness of the film (top horizontal axis) and the inverse of the thickness (bottom horizontal axis) with zero corresponding to the bulk result.
The two upper curves represent the energy difference between the second conduction band and the last but one valence band ($E^{CBM+1}-E^{VBM-1}$) obtained with TB based on DFT in the local density approximation (LDA-TB, red squares) and on QS$GW$ (purple circles). This energy difference, defined by the quantum-well states at the $\overline{\rm{M}}$ point (see Fig.~\ref{fig2-200}), should tend to the bulk band gap value for an infinite thickness.
The real bulk values, \ie\ obtained in an \emph{infinite} (3D-periodic) LDA or QS$GW$ calculation, are shown as open symbols, and the thickness-dependent values indeed converge towards these bulk values, as expected. The cyan small circles represent the evolution of the actual band gap, i.e., the gap between the first conduction band and the last valence band, obtained with QS$GW$-TB.

If the surface states of the film in Fig.~\ref{fig2-200} were the result of a non-trivial topology, this band gap would not tend to zero but to the bulk band gap. In other words, the separation between quantum-well states would get smaller as the thickness increases, but both states would remain within the bulk projection, so the two curves (cyan and purple) would approach each other and, at infinite thickness, both would tend to the bulk value (see also Fig.~S2 in the Supplemental Material~\cite{suppmat2021} for a schematic visualization of this). 
This is not the case and, as seen in Fig.~\ref{fig3-thick}, the band gap indeed tends to zero.
From our analysis in Ref.~\cite{aguilera2015}, we know that bulk Bi is a trivial semimetal when calculated with QS$GW$. From Fig.~\ref{fig3-thick}, 
we know that the surface states of films of any thickness are not the result of a non-trivial $Z_2$ topology. This proves that the surface states in Fig.~\ref{fig2-200} are trivial in spite of their non-trivial appearance. 

In the following, we discuss the origin of the seemingly non-trivial character. 
It has already been discussed in the literature, especially in Ref.~\cite{ishida2016}, that the surface states of Bi have a very long decay length into the bulk. 
From topological insulators, we know that a long decay length of the surface states can give rise to a ``crosstalk'' effect: If 
the wavefunctions of the surface states at both surfaces of a film decay into the bulk deep enough to overlap with each other, the two 
surface states might hybridize, which can open a gap where there should otherwise be a degeneracy. 
The calculations of the decay length of the surface states of Bi were so far performed with DFT. 
The most recent predictions of this length were 24 BLs (Ref.~\cite{ishida2016}) and 21 BLs (Ref.~\cite{chang2019}). This
means that films of at least 42 BLs need to be studied to find a degeneracy at the $\overline{\rm{M}}$ point as in Fig.~\ref{figscheme}(a). For thinner films, one should observe a hybridization gap between the surface states leading to a situation more similar to that in Fig.~\ref{figscheme}(b).  
But the experimental works from Refs.~\cite{yao2016} and \cite{ito2016} observed such behavior for much thicker samples, typically considered bulk-like (80 BLs and 202 BLs, respectively), which led the authors to conclude in the end that the topology is non-trivial. 

However, it should be kept in mind that the decay length depends inversely on the size of the band gap~\cite{linder2009}. 
In the case of DFT with the standard approximations used in 
Refs.~\cite{chang2019} and~\cite{ishida2016}, 
inverted band gaps like the one of bismuth at the L point are overestimated~\cite{aguilera2015}. In particular, for Bi, the DFT band gap at L is 6 times larger than the experimental one. 
This leads to an underestimated decay length of the surface states. We have performed calculations with the QS$GW$-TB method for films of thicknesses up to 2000 BLs (approximately 800~nm). Even for a film of 1000~BLs, the two surface states are not yet degenerate at the $\overline{\rm{M}}$-point   
and there is still a small band gap ($\sim$0.3~meV) between them. For 2000~BLs, the two states can already be considered degenerate at $\overline{\rm{M}}$, 
with a gap between them of only 0.005~meV.
This means that when calculations of the decay length are performed 
with a method that can predict the band gap accurately, the decay length of the surface states of Bi must be larger than 500 BLs. 
This can explain why the samples of Refs.~\cite{yao2016} and \cite{ito2016} present a connectivity of the surface states with 
bulk valence and conduction bands without having to assume a non-trivial topology. Actually, the fact that a 200 BLs film 
of Bi cannot be considered bulk-like should not be surprising, given that one can experimentally clearly observe well-defined quantum-well states close 
to $\overline{\rm{M}}$ (see Fig.~3 of Ref.~\cite{ito2016}). 

It should also be noted that in Refs.~\cite{ohtsubo2013,ohtsubo2016}
the possibility of an interaction between top and bottom surface was already mentioned. However, the authors interpreted
that the ``inter-surface interaction would re-invert the bulk bands at L'' 
(projected to $\overline{\rm{M}}$), resulting in a change of topology. As we have discussed above and as 
is supported by the measurements of Ref.~\cite{ito2016}, there is no evidence of such effect 
as a function of 
thickness and therefore, no change in topology occurs caused by the thickness or by the surface-surface interaction.
So, although it may be counterintuitive, the spectra in Refs.~\cite{ohtsubo2013,ohtsubo2016,yao2016,ito2016} are perfectly consistent with a trivial situation.
In the case of our calculation in Fig.~\ref{fig2-200}, we are indeed certain that the system is trivial and the ``topological-looking'' connectivity is a consequence of crosstalk and not of a non-trivial topology. 
We should mention here that this picture does not change if the two surfaces are slightly inequivalent due to the presence of a substrate on one side. Although the dispersion of the upper and lower surface states changes slightly, the connectivity near the $\overline{\rm{M}}$ point remains almost unchanged (see Fig.~S3 of the Supplemental Material~\cite{suppmat2021}).

\begin{figure}
\begin{center}
\includegraphics[angle=0,width=8.4cm]{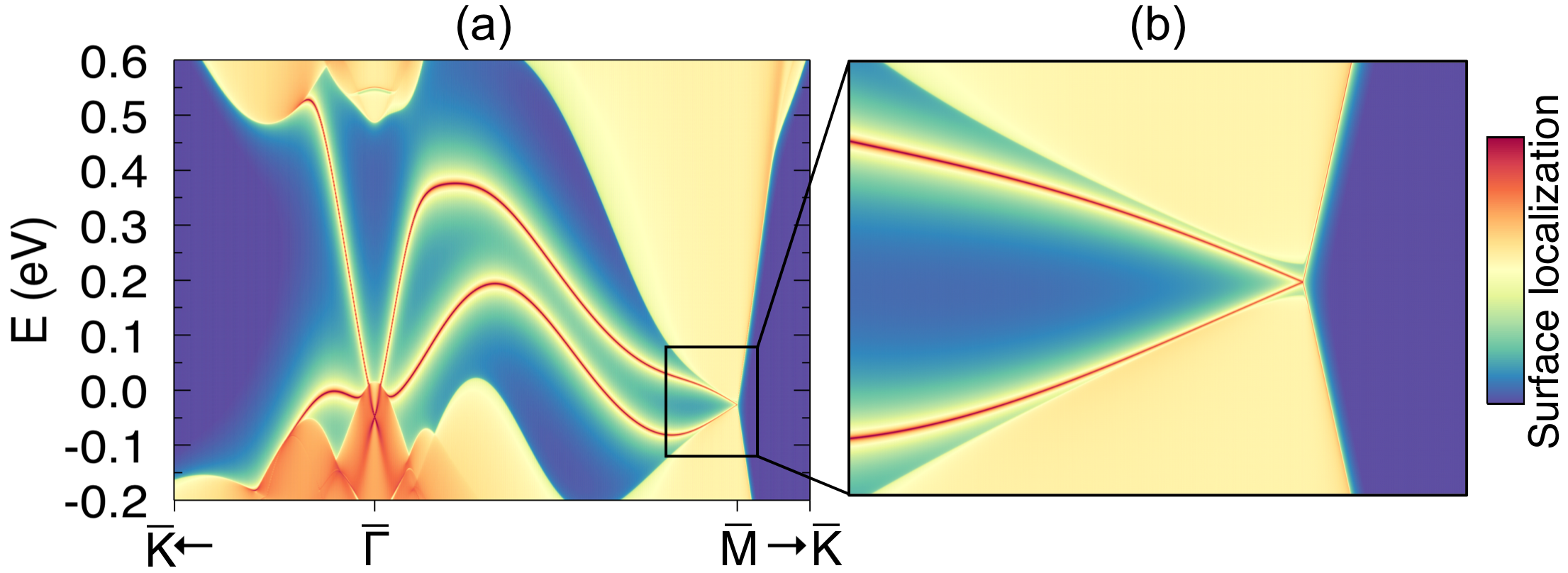}
\caption{\label{semiinf}Spectral function of a semi-infinite calculation of Bi obtained with the iterative Green's function method and a TB Hamiltonian based on QS$GW$ calculations. (b) displays a magnification of (a) around the $\overline{\rm{M}}$ point.}
\end{center}
\end{figure}

To further support our explanation based on a crosstalk effect, we performed a semi-infinite calculation
with the iterative Green's function method~\cite{lopezsancho1984} as implemented in {\sc WannierTools}{}~\cite{WU2017}.
For that, we used again the TB Hamiltonian with parameters obtained using QS$GW$ calculations. The results are shown in Fig.~\ref{semiinf} and prove that in the limit of infinite thickness (where there is 
obviously no crosstalk), the surface states behave as expected according to the trivial fashion of Fig.~\ref{figscheme}(a).  
The surface states are degenerate at $\overline{\rm{M}}$ and they do not connect separately to the bulk valence and conduction band.

\begin{figure}
\begin{center}
\includegraphics[angle=0,width=8.2cm]{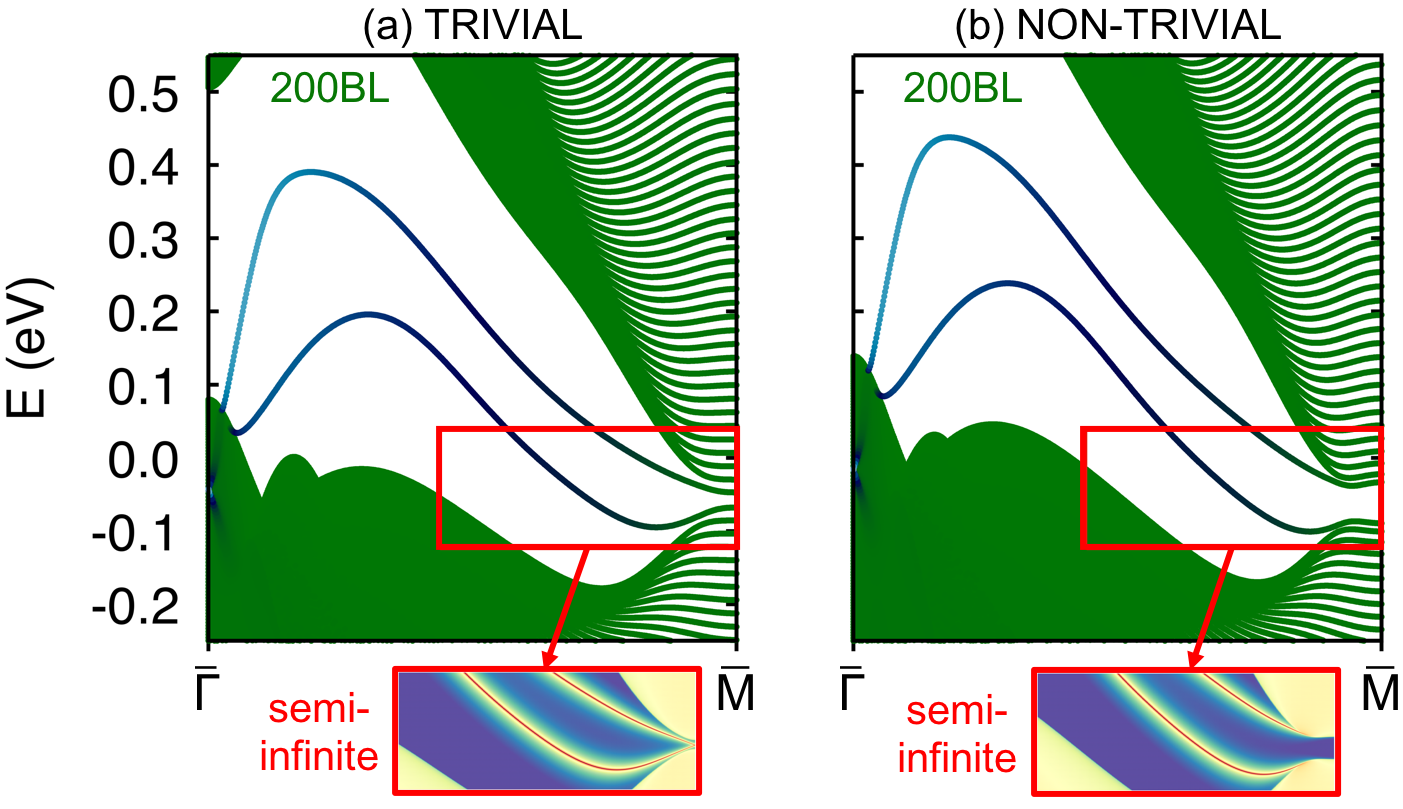}
\caption{Band structure of 200-BL films (top panels) of Bi and spectral function of a semi-infinite calculation (bottom panels). Results are obtained with a TB Hamiltonian based on one-shot $GW$ calculations. The films have been strained to represent two cases very close to the topological phase transition (see main text). As in Figs.~\ref{fig2-200} and \ref{semiinf}, the color schemes represent the localization on the surface. \label{figstrained}}
\end{center}
\end{figure}

Making use of the fact that Bi undergoes a topological phase transition under strain~\cite{aguilera2015}, we can now also compare calculations for a trivial and a non-trivial system. From Fig.~3(a) of Ref.~\cite{aguilera2015}, we can choose two strained states of Bi that have approximately the same band gap but opposite topology. 
In Fig.~\ref{figstrained}, we show results for two systems under volume-conserving compressive strain [0.5\% and 1.0\% in-plane strain in panels (a) and (b), respectively]. These two strain values were chosen close to the phase transition on the trivial and non-trivial sides. 
The main (top) panels show band structures of 200-BLs of Bi obtained with a TB Hamiltonian based on the one-shot $GW$ method including off-diagonal elements of the self-energy matrix~\cite{aguilera2013}. The surface states of the two 200-BL films in Fig.~\ref{figstrained} look qualitatively the same and this may lead to the conclusion that these films are both non-trivial. But when calculations are performed in a semi-infinite geometry (bottom panels), the differences become evident and one can now clearly distinguish between the two phases: the trivial one in Fig.~\ref{figstrained}(a), with degenerate surface states at $\overline{\rm{M}}$; and the topological one in Fig.~\ref{figstrained}(b), with one surface state connecting to the valence band and the other one to the conduction band.
Figure~\ref{figstrained} thus proves that one cannot distinguish between a trivial and a non-trivial phase of Bi only by looking at the connectivity of the surface states with the valence and conduction states unless the bulk limit has really been reached, which is not the case for 200~BLs. 

To summarize, there is no inconsistency between theory and experiment about the surface states of Bi films. 
The surface states predicted by the advanced QS$GW$ method 
are in very good agreement with the ARPES spectra. The theory predicts that bulk Bi is trivial. The surface states of Bi films might look non-trivial because of a crosstalk effect due to the extremely large 
decay length of the surface states into the bulk. 
We have discussed both finite films and the thermodynamic limit of semi-infinite systems and we have shown that bismuth starts behaving like bulk only after at least 1000~BLs.
All this shows that one cannot draw conclusions about the topology of a given sample just by observing the connectivity of the surface states with the valence and conduction bands, and the surface states of Bi(111), claimed to be topological in Refs.~\cite{ohtsubo2013,ohtsubo2016,yao2016,ito2016}, are fully consistent with a trivial phase. This resolves the long-standing controversy about the $Z_2$ character of Bi.
We would like to point out that according to Ref.~\cite{aguilera2015}, the strain needed to drive Bi into a topological 
semimetal is very small, and therefore, depending on the doping, choice of substrate, or growth conditions, one cannot rule out that some of the experimentally measured samples might 
actually be in a non-trivial state~\footnote{For example, in Ref.~\cite{nayak2019}, they attributed the non-trivial topology observed to the 
strain induced in the vicinity of a dislocation. They concluded that their samples consisted of a non-trivial cylindrical 
volume (around the dislocation) embedded in a $Z_2$ trivial bulk.}.

To conclude, irrespective of the possible existence of both trivial and non-trivial Bi samples,
this work has shown that the observation of a seemingly non-trivial surface-bulk connectivity can be deceptive in the case of a film geometry. 
If the film is not thick enough, a crosstalk effect of interacting surface wavefunctions, penetrating deep into the sample from both sides of the film, can explain the observation equally well, and no conclusive answer as to whether the respective bulk material is trivial or not can be given. Therefore, predictions from DFT calculations have to be interpreted with care as an overestimation of the band gap of a few meV can lead to a grossly underestimated penetration depth of the surface states.
Since deeply penetrating surface states are not only characteristic of Bi but also common in topological insulators with their narrow band gaps, the crosstalk modification of the band connectivity is a general issue in topological matter. 
In fact, our conclusions are also applicable to the edge states of ribbons of 2-dimensional topological insulators, \ie\ quantum spin Hall as well as Chern insulators, if the ribbon is not wide enough to decouple the two edges.

\section*{Acknowledgments}
S.B.\ gratefully acknowledges financial support from Deutsche For\-schungs\-gemeinschaft 
(DFG) through the  Collaborative  Research Centers SFB 1238 (Project C01) and  H.-J. K.\ from the Alexander von Humboldt Foundation. 
We acknowledge the computing time granted through JARA-HPC on the supercomputer JURECA at Forschungszentrum J{\"{u}}lich.

\end{document}


\title{Supplemental Material for\\ $Z_2$ topology of bismuth}

\author{Irene Aguilera}
\author{Hyun-Jung Kim}
\author{Christoph Friedrich}
\author{Gustav Bihlmayer}
\author{Stefan Bl\"{u}gel}
\affiliation{Peter Gr{\"{u}}nberg Institute and Institute for Advanced Simulation, Forschungszentrum J{\"{u}}lich and JARA, 52425 J{\"{u}}lich, Germany}

\date{\today}

\maketitle

\begin{figure}[!h]
\begin{center}
\includegraphics[angle=0,width=16cm]{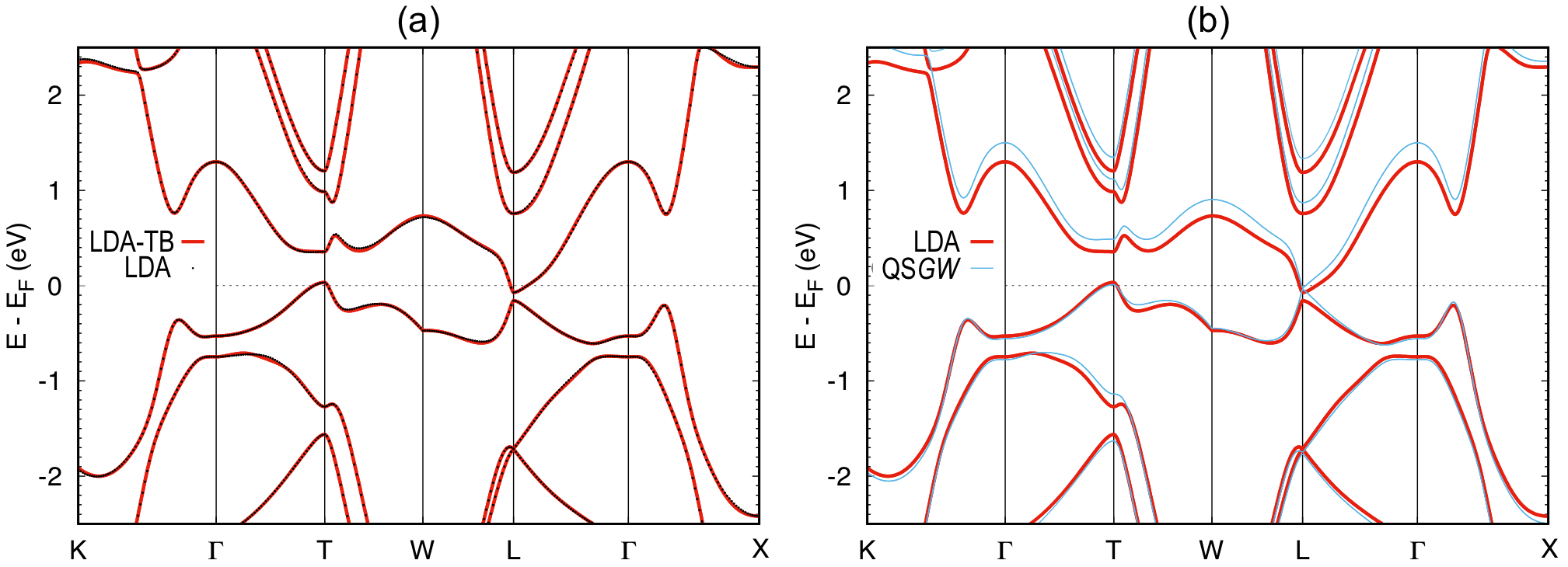}
\caption{\label{figbandsbulk} Band structure of bulk Bi. (a) Comparison between the explicitly calculated DFT-LDA band structure (labelled LDA) and the one interpolated with the Wannier interpolation method that is used to construct the tight-binding Hamiltonians used in this work (labelled LDA-TB). (b) Comparison of the band structures obtained with a Wannier interpolation based on an LDA and a QS$GW$ calculation.}
\end{center}
\end{figure}

Figure~\ref{figbandsbulk} shows band structures of \emph{bulk} Bi as support to the \emph{film} results of the main manuscript. 
These band structures have been previously published in Ref.~\cite{aguilera2015} and the details of the calculations are given in that publication.

\newpage 

\begin{figure}[!h]
\begin{center}
\includegraphics[angle=0,width=11cm]{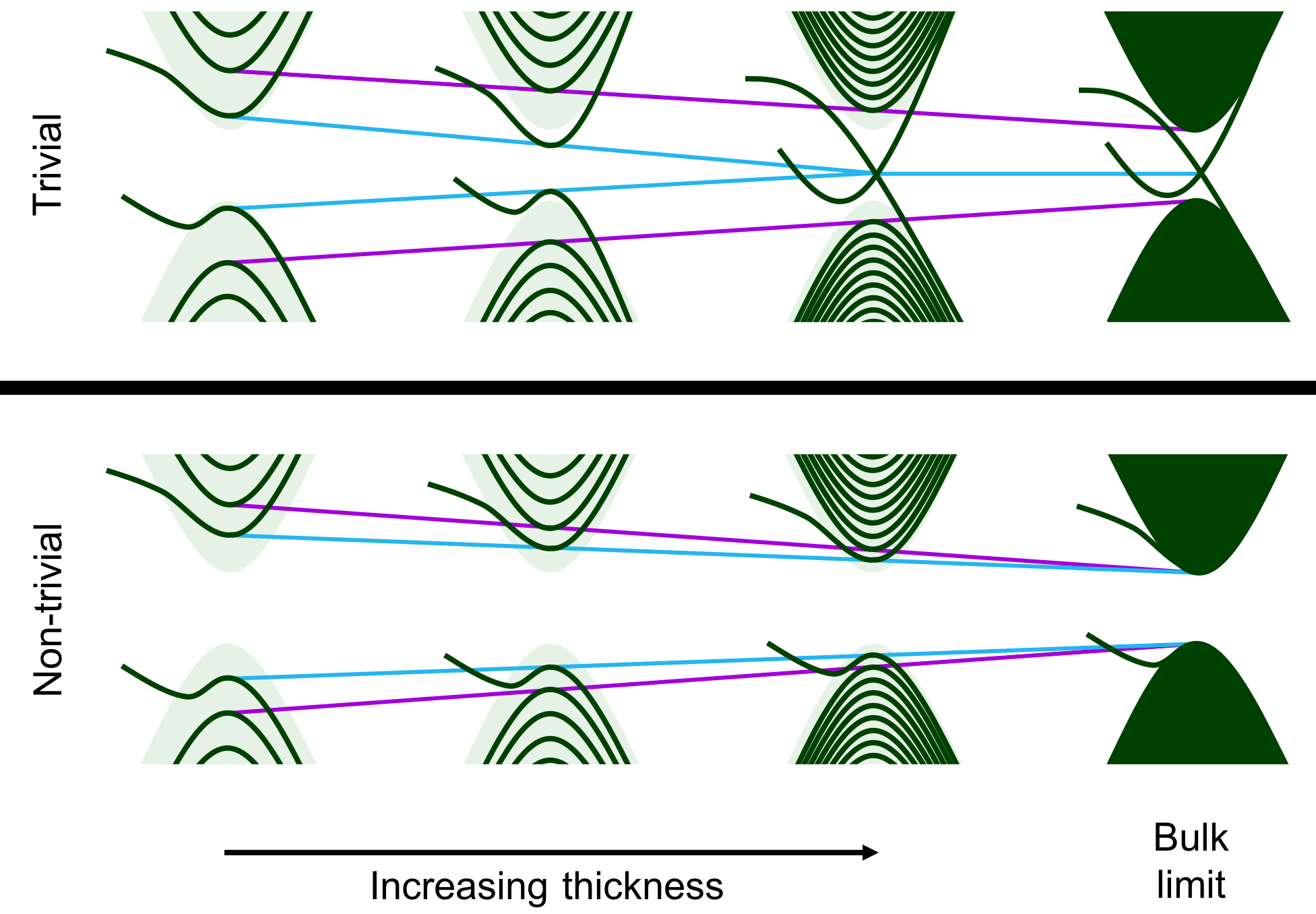}
\caption{\label{figscheme} Schematic representation of the 
two possible scenarios for the band gap at the $\overline{\rm{M}}$ point as a function of the thickness of the film. 
Top panels: In the trivial case, the surface states  
connect to the bulk (quantum wells) when the films are not thick enough as a result of crosstalk. 
For a certain thickness, the surface states clearly separate from the bulk continuum (light green background)
and for thicknesses above the critical one (when there is no crosstalk anymore), 
the two surface states are degenerate at $\overline{\rm{M}}$. 
Bottom panels: In the non-trivial case, the surface states 
at $\overline{\rm{M}}$ will remain within the projection of 
the bulk continuum for any thickness and both curves
discussed in Fig.~3 (cyan and purple) would tend to the band gap value
of the infinite bulk.}
\end{center}
\end{figure}

Figure~\ref{figscheme} supports the discussion about Fig.~3 of the main manuscript that proves that the surface states observed in Fig.~2 are trivial. If they were not, we would observe a dependence of the values of Fig.~3 as corresponding to the picture in the bottom panels of Fig.~\ref{figscheme}. But instead, in Fig.~3 we show that the band gaps of the films tend to zero, while the energy difference $E^{CBM+1}-E^{VBM-1}$ tends to the real (infinite) bulk value, as in the top panels of Fig.~\ref{figscheme}.

\newpage

\begin{figure}[!h]
\begin{center}
\includegraphics[angle=0,width=16cm]{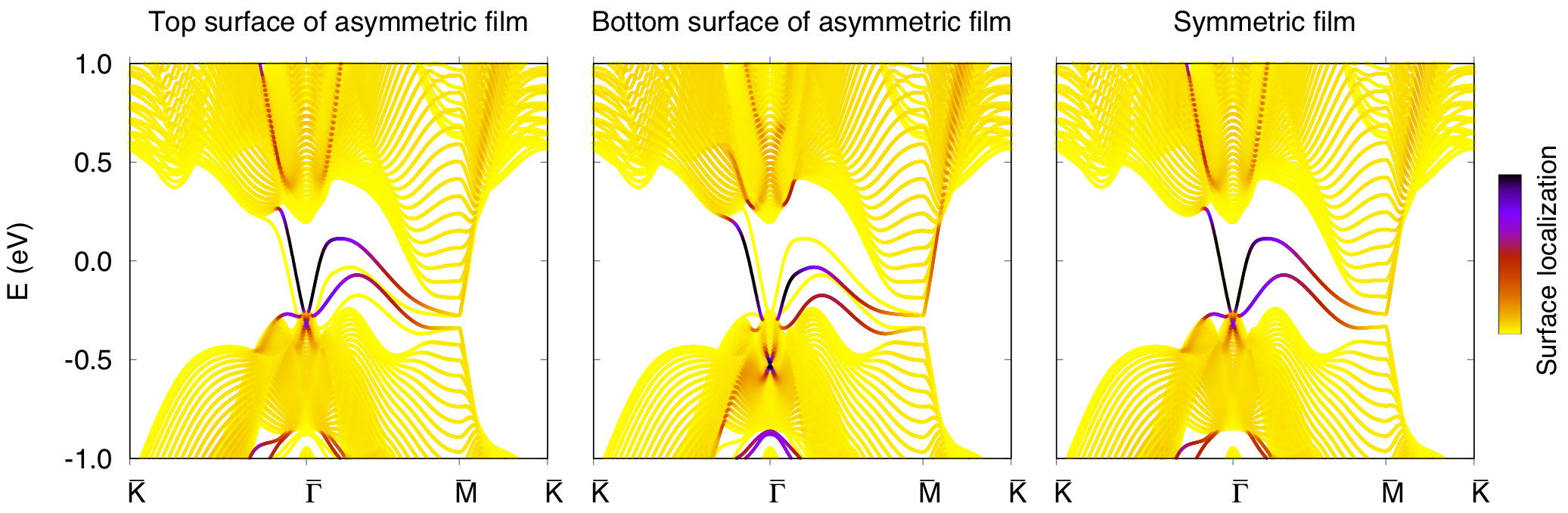}
\caption{\label{figasymm} Band structure of a 50-BL film of Bi(111) along some high-symmetry lines in the two-dimensional Brillouin zone obtained with the QS$GW$-TB method. The color scheme represents the degree of localization of a state at the surface. The left and middle panels show the two inequivalent surfaces of an asymmetric film, whereas the right panel shows the symmetric film as a reference.}
\end{center}
\end{figure}

To simulate an asymmetric film (that might occur in practice due to, e.g., the presence of a substrate), we have modified the tight-binding parameters corresponding to the bottom surface. For that, we have rescaled the Hamiltonian matrix elements associated to the atoms of the bottom bilayer by reducing them by 0.03\%. The results for a film of 50 bilayers with inequivalent surfaces is shown in Fig.~\ref{figasymm}. Although the dispersion of the top and bottom surface states is slightly different, the connectivity near the $\overline{\rm{M}}$ point (and the band gap) remains almost unchanged.

%

%